\documentclass[useAMS,usenatbib]{mn2e}

\usepackage{amsmath}
\usepackage{amssymb}
\usepackage{graphicx}

\usepackage{natbib}

\usepackage[draft]{hyperref}

\newcommand{\dd}{\,\mathrm{d}}

\newcommand{\fofrfour}{fofr4\ }
\newcommand{\fofrfive}{fofr5\ }
\newcommand{\fofrsix}{fofr6\ }
\newcommand{\symmA}{symm\_A\ }
\newcommand{\symmB}{symm\_B\ }
\newcommand{\symmC}{symm\_C\ }
\newcommand{\symmD}{symm\_D\ }

\newcommand{\fig}[1]{Fig.~\ref{#1}}
\newcommand{\figs}[2]{Figs.~\ref{#1}--\ref{#2}}

\newcommand\apj{\rmfamily{ApJ}}%
\newcommand\apjl{\rmfamily{ApJ}}%
\newcommand\aap{\rmfamily{A\&A}}%
\newcommand\jcap{\rmfamily{J. Cosmology Astropart. Phys.}}%
\newcommand\mnras{\rmfamily{MNRAS}}%
\newcommand\prd{\rmfamily{Phys.~Rev.~D}}%
\newcommand\prl{\rmfamily{Phys.~Rev.~Lett.}}%
\newcommand\nat{\rmfamily{Nature}}%
\newcommand\physrep{\rmfamily{Phys.~Rep.}}%
\title[Halo velocity profiles in screened modified gravity theories]{Halo velocity profiles in screened modified gravity theories}
\author[M. Gronke, C. Llinares,  D. F. Mota and H. A. Winther]{M. Gronke$^{1}$\thanks{E-mail:
maxbg@astro.uio.no},  C. Llinares$^{1}$, D. F. Mota$^{1}$ and H. A. Winther$^{2}$\\
$^{1}$Institute of Theoretical Astrophysics, University of Oslo, Postboks 1029, 0315 Oslo, Norway\\
$^{2}$Astrophysics, University of Oxford, DWB, Keble Road, Oxford, OX1 3RH, UK}

\begin{document}

\date{Preprint version from \today}

\pagerange{\pageref{firstpage}--\pageref{lastpage}} \pubyear{2014}

\maketitle

\label{firstpage}

\begin{abstract}
Screened modified gravity predicts potentially large signatures in the peculiar velocity field that makes it an interesting probe to test gravity on cosmological scales. We investigate the signatures induced by the Symmetron and a Chameleon $f(R)$ model in the peculiar velocity field using $N$-body simulations. By studying fifth force and halo velocity profiles we identify three general categories of effects found in screened modified gravity models: a fully screened regime where we recover $\Lambda$CDM to high precision, an unscreened regime where the fifth force is in full operation, and, a partially screened regime where screening occurs in the inner part of a halo, but the fifth force is active at larger radii. These three regimes can be pointed out very clearly by analyzing the deviation in the maximum cluster velocity. Observationally, the partially screened regime is of particular interest since an uniform increase of the gravitational force -- as present in the unscreened regime -- is degenerate with the (dynamical) halo mass estimate, and, thus, hard to detect. 
\end{abstract}


\section{Introduction}
The accelerated expansion of the Universe is one of the most challenging problems of modern cosmology. So far, this expansion can be well modelled by a cosmological constant in General Relativity (GR) resulting in the $\Lambda$CDM model. Alternatives to $\Lambda$CDM are numerous \citep[see, e.g.,][]{Amendola2010,Clifton2012}, but numerous are also the problems they have to struggle with: some of them are plagued by theoretical instabilities and all of them have to face observational constraints which often requires fine-tuning of the model parameters. One particularly simple extension of GR is the inclusion of a single scalar field $\varphi$ to the Einstein-Hilbert Lagrangian which sources GR. However, when coupled to matter, the scalar field gives rise to an additional gravitational force -- often called a fifth force \citep[see e.g. ][]{Amendola2013,Mota2006,Hellwing2013}. This fifth force can be quantified by $\gamma\equiv |{\bf F}_{\rm Fifth}|/|{\bf F}_{\rm N}|$ where $F_{\ rm N}$ is the `standard' Newtonian gravitational force that we get in the weak-field limit of GR. Several experiments \citep[see e.g.][]{Adelberger2002cls..conf....9A,Berotti2003Natur.425..374B,Williams2004PhRvL..93z1101W,Will2006LRR.....9....3W} have constrained $\gamma \ll 1$ on Earth and in the solar-system. This leaves two kind of explanations: either GR is correct on all scales, i.e., $\gamma = 0$, or $\gamma$ is not a constant but instead varies in space (and time).

Models where the latter is the case (i.e. where $\gamma$ depends on the local environment) are dubbed \textit{screened modified gravity models} since the fifth force is screened in high-density environments. Five types of screening mechanisms are currently discussed in the literature \citep[see, e.g.,][]{Khoury2010,Joyce2014}:
\begin{enumerate}
\item the Vainshtein Mechanism \citep{Vainshtein1972PhLB...39..393V}, where the fifth force is suppressed by the large kinetic terms appearing in high-density regions,
\item Chameleon screening \citep{Khourya,Khoury} in which the mass of the scalar field (and, thus, the range of the fifth force) depends on the matter density/the space-time curvature,
\item the Symmetron model \citep{Hinterbichler,Hinterbichlera}, where the coupling strength is the density-dependent quantity, and, relatively new
\item the disformal screening mechanism \citep{Koivisto2012PhRvL.109x1102K}, as well as,
\item D-BIonic screening \citep{Burrage2014arXiv1403.6120B}.
\end{enumerate}
Out of these, the nonlinear effects arising from mechanisms (i)--(iii) are studied numerically using cosmological $N$-body simulations in which the full scalar field evolution and its effect on structure formation is modelled \citep[e.g.,][]{Brax2012,Brax2013a,Davis2011,Li2013JCAP...11..012L,paper1,paper2,Llinares2014a}.
In this paper, we will focus on the latter two of them, leaving out the characteristically distinct Vainshtein screening which plays a role in certain DGP and Galileon models \citep{DGP2000PhLB..485..208D, Nicolis2009PhRvD..79f4036N,Chow2009PhRvD..80b4037C}.

Generally, the here studied screened modified gravity models behave on the background level like $\Lambda$CDM \citep{2012PhRvD..86d4015B}. In addition, in regions where small-scale effect can be most easily studied (i.e., the solar system), the fifth force is screened\footnote{An exception of this is potentially the search for a varying fine structure constant $\alpha$. See, e.g., \citet{Flambaum2008EPJST.163..159F}.}. Therefore, intermediate scales -- that is galaxy- and cluster-size scales -- are of particular interest. For the Symmetron as well as for Chameleon-type models, several potential observables present on such length-scales have been studied previously \citep[e.g.][]{Brax2013a,Brax2012,Winther2011,Llinares2013b,Gronke2014, 2012PhRvL.109e1301L, 2014arXiv1410.2857L}. 

In this paper, we also want to focus on these length-scales and study the dynamics of clusters of galaxies in the Symmetron and the Chameleon-$f(R)$ model. We are interested in studying the impact that modified gravity has on the properties of the dark matter velocity field. The fact that modified gravity can have an influence on this observable was previously discussed, e.g.,  by \citet{2009ApJ...695L.145L} or \citet{2012ApJ...747...45L} in the context of the collisional velocity of dark matter halos. Here, we focus on the profiles of the velocity field of clusters of galaxies. We expect these profiles to be directly affected by the enhanced gravitational force and thus, to be able to act as a useful observable that could help to distinguish between GR and modified gravity.

The paper is structured as follows: In Sec.~\ref{sec:methods}, we introduce the screened modified gravity models employed, namely the Symmetron and the Hu-Sawicky $f(R)$ gravity and explain our method used for the analysis. We present our results in Sec.~\ref{sec:results} and discuss them in Sec.~\ref{sec:discussion}. Finally, we conclude in Sec.~\ref{sec:conclusions}.

\section{Methods}
\label{sec:methods}
In the following, the standard notation of normalizing background densities to the critical density defined as $\rho_c = 3 H^2 M_{\rm Pl}^2$ and denoting them with $\Omega$ is applied. Here, $M_{\rm Pl}$ and $H$ are the reduced Planck mass which is defined as $M_{\rm Pl}^{-2} \equiv 8 \pi G$ and the Hubble parameter $H$, respectively. Also note that a subscript $0$ denotes the values today.

\subsection{Modified Gravity Models}
\label{sec:mod_grav_models}
The characteristic quantity of screened modified gravity models is the fifth force, ${\bf F_{\rm Fifth}}$, which is an additional contribution to the (Newtonian) gravitational force and varies in space and time. In particular, screened modified gravity models are constructed so that $|{\bf F}_{\rm Fifth}| \ll |{\bf F}_{\rm N}|$ in the solar system to evade constraints from local gravity experiments.

This variable gravitational force arises from an additional scalar field $\varphi$ added to the Einstein-Hilbert action
\begin{align}
S =& \int\dd^4x \sqrt{-g}\left(\frac{M_{\mathrm{Pl}}^2}{2}R-\frac{1}{2}\nabla_\mu\varphi\nabla^\mu\varphi-V(\varphi)\right) \nonumber\\ 
\,&+ \mathcal{S}_m(\psi^{(i)},\tilde g_{\mu\nu})
\label{eq:action}
\end{align}
where the matter fields $\psi^{(i)}$ couple to the Jordan frame metric $\tilde g_{\mu\nu}\equiv A^2(\varphi)g_{\mu\nu}$ and $R$ is the Ricci scalar.

Taking the variation of Eq.~\eqref{eq:action} with respect to the metric $g_{\mu\nu}=-(1+2\Phi)\delta_{\mu 0}\delta_{\nu 0}+ a^2(1-2\Phi)\delta_{\mu i}\delta_\nu^i$ in the Newtonian gauge gives us the Einstein equations
\begin{equation}
G^{\mu\nu} = A(\varphi)T^{\mu\nu}_{m} + T_\varphi^{\mu\nu}
\end{equation}
where $T^{\mu\nu}_{m}$ is the stress-energy tensor the matter fields and $T_\varphi^{\mu\nu} = \varphi_{,\mu}\varphi_{,\nu} - g_{\mu\nu}\left(\frac{1}{2}\nabla_{\alpha}\varphi\nabla^{\alpha}\varphi + V(\varphi)\right)$ is the stress-energy tensor of the scalar field. By taking $T^{\mu\nu}_{m}$ to be that of non-interacting particles
\begin{equation}
T^{\mu\nu}_{m}({\bf y}) = \sum_{i}\frac{m_i}{\sqrt{-g}}\delta({\bf y-x_i})\dot{x}^{\mu}_i\dot{x}^{\nu}_i
\end{equation}
and using the conservation equation $\nabla_{\mu}(A(\varphi)T_m^{\mu\nu} + T_\varphi^{\mu\nu}) = \nabla_{\mu}G^{\mu\nu} = 0$ we obtain the geodesic equation for the matter particles\footnote{Note, that here the quasi-static limit is applied. For a full discussion on the impact of this approximation in the Symmetron and $f(R)$ cases see \citet{2013PhRvL.110p1101L, 2013arXiv1310.3266N, Llinares2014b, Bose2014arXiv1411.6128B}.}
\begin{equation}
  \mathbf{\ddot x}+2H\mathbf{\dot x} +\frac{1}{a^2}\vec\nabla\left(\Phi+\log A\right) = 0.
  \label{eq:geodesic_general}
\end{equation}
Here, $\Phi$ is usually associated with the Newtonian gravitational potential, $a(t)$ is the scale factor and $H\equiv \dot a/a$ is the Hubble parameter. From the last term, it is clear that the fifth force on a test particle with mass $m$ is given by
\begin{equation}
{\bf F_{\rm Fifth}} = -\frac{m}{a^2}\vec\nabla\log A.
\label{eq:fifth_force_general}
\end{equation}

The evolution of the scalar field, on the other hand, can be obtained from $\delta S/\delta\varphi = 0$ and reads 
\begin{equation}
\square\varphi = \frac{dV_{\rm eff}}{d\varphi}
\end{equation}
where $V_{\rm eff}$ is the effective potential. In dust dominated regions we have
\begin{equation}
  V_{\text{eff}}=V(\varphi)+\left(A(\varphi)-1\right)\rho_m
  \label{eq:effective_potential}
\end{equation}
where $\rho_m$ is the matter density. This term is of special interest since the range of the fifth force is given by
\begin{equation}
\lambda_{\rm Fifth} = \left(\frac{\dd^2 V_{\rm eff}}{\dd \varphi^2}\big|_{\rm min}\right)^{-1/2}
\label{eq:fifth_range_general}
\end{equation}
where the derivative is evaluated at the (space-time dependent) minimum of the effective potential.

In the following, we will briefly describe the two screened modified gravity models utilized in this paper. For a more complete description of the models see the original papers \citet{Hinterbichler} and \citet{Hu2007}.

\subsubsection{Symmetron}
In the Symmetron model \citep{Hinterbichler,Hinterbichlera}, the coupling function as well as the potential are symmetric around $\varphi = 0$ and are given by
\begin{align}
A(\varphi) &= 1 + \frac{\varphi^2}{M^2} \\
V(\varphi) &= -\frac{1}{2}\mu^2\varphi^2+\frac{1}{4}\lambda\varphi^4\;.
\end{align}
With these definitions, the effective potential Eq.~\eqref{eq:effective_potential} becomes 
\begin{equation}
V_{\mathrm{eff}} = \frac{1}{2}\left(\frac{\rho_m}{M^2}-\mu^2\right)\varphi^2+\frac{1}{4}\varphi^4
\end{equation}
with a minimum at $\varphi_{\rm hd}=0$ for $\rho_m\ge \rho_{\rm ssb}\equiv \mu^2 M^2$ and two minima at $\varphi_{{\rm ld},\pm}=\pm \sqrt{\mu^2 - \rho_m / M^2}$ otherwise.
Consequently, the fifth force ${\bf F}_{\rm Fifth} \propto -\varphi\vec\nabla\varphi + \mathcal{O}(\varphi^3/M^3)$ (from Eq.~\eqref{eq:fifth_force_general}) is screened in the former case.

As previously in \citet{Gronke2014}, we redefine the model parameters $(M, \mu, \lambda)$ to the range of the fifth force in vacuum
\begin{equation}
  L = \frac{1}{\sqrt{2}\mu},
\end{equation}
the scale factor at time of symmetry breaking (i.e., when the average matter density is equal to $\rho_{\rm ssb}$)
\begin{equation}
a_{\rm ssb}^3 = \frac{\Omega_{m0}\rho_{c0}}{\mu^2 M^2},
\end{equation}
and, the coupling strength
\begin{equation}
  \beta = \frac{\mu M_{\rm Pl}}{M^2 \sqrt{\lambda}}.  
\end{equation}

\begin{table}
  \caption{Model parameters of the different simulation runs.}
  \centering
  \begin{tabular}{@{}lrr@{}}
    \hline \hline
    Name & $|f_{R0}|$ & $n$ \\
    \hline
    \fofrfour & $10^{-4}$ & $1$ \\
    \fofrfive & $10^{-5}$ & $1$ \\
    \fofrsix & $10^{-6}$ & $1$ \\
    \hline
  \end{tabular}

  \begin{tabular}{@{}lrrc@{}}
    \hline \hline
    Name & $a_{\rm ssb}$ & $\beta$ & $L$ \\
         &          &         & (Mpc $h^{-1}$) \\
    \hline
    \symmA & $0.50$ & $1.0$ & $1.0$ \\
    \symmB & $0.33$ & $1.0$ & $1.0$ \\
    \symmC & $0.50$  & $2.0$ & $1.0$ \\
    \symmD & $0.25$ & $1.0$ & $1.0$ \\
    \hline
  \end{tabular}
\label{tab:run_params}
\end{table}

\begin{figure*}
\centering
\includegraphics[width=0.45\linewidth]{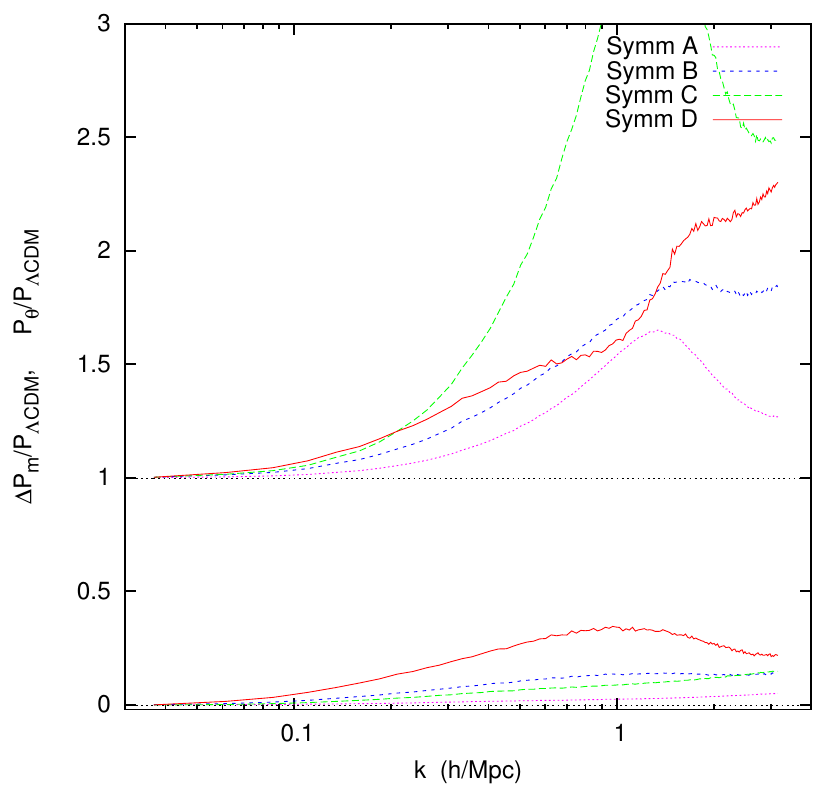}
\includegraphics[width=0.45\linewidth]{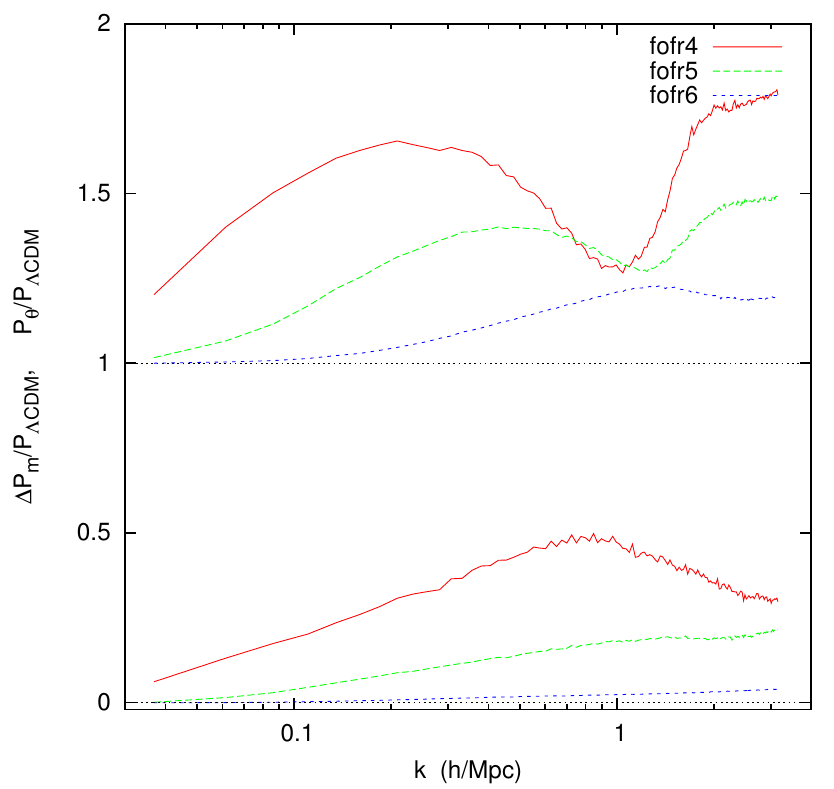}
\caption{The fractional difference of the velocity divergence (above) and matter (below) power-spectra with respect to $\Lambda$CDM for the Symmetron (left) and $f(R)$ (right) simulations.}
\label{fig:pofkv}
\end{figure*}

\subsubsection{Hu-Sawicky $f(R)$ gravity}
$f(R)$-models are a group of modified gravity theories in which to the single Ricci scalar $R$ in the Einstein-Hilbert action a generic function $f(R)$ is added. Here, we consider the Hu-Sawicky model \citep{Hu2007} which incorporates a Chameleon screening. 

However, through a conformal transformation $\tilde g_{\mu\nu}=A^2(\varphi)g_{\mu\nu}$ where $A(\varphi) = \mathrm{e}^{\beta\varphi/M_{\rm Pl}}$ with $\beta = 1/\sqrt{6}$ one can bring this $f(R)$-action on the form of Eq.~\eqref{eq:action} \citep[see, for example,][]{2006CQGra..23.5117S}. The potential becomes
\begin{align}
  V(\varphi) =& M_{\rm Pl}^2\frac{f_R R-f}{2(1+f_R)^2},
\end{align}
where $f_R \equiv \dd f/\dd R = \exp(-2\beta\varphi/M_{\rm Pl})-1\approx -\frac{2 \beta \varphi}{M_{\rm Pl}}$. Using this relation, we can rewrite the potential as
\begin{equation}
V(\varphi) = \rho_\Lambda - \frac{n + 1}{2 n} M^2_{\rm Pl} R_0 |f_{R0}| \left(\frac{2 \beta \varphi}{M_{\rm Pl} |f_{R0}|}\right)^{\frac{n}{n+1}}.
\nonumber
\end{equation}
Here, $\rho_\Lambda = 3H_0^2 M_{\rm Pl}^2\Omega_{\Lambda}$ is an effective cosmological constant and $R_{0} = 3H_0^2 (\Omega_{m0}+4\Omega_{\Lambda0})$ is the background curvature today.

This implies, the strength of the fifth force (see Eq.~\eqref{eq:fifth_force_general}) is proportional to $\nabla\varphi$ which is not necessarily small in high density environments. However, the range of the force today given by Eq.~\eqref{eq:fifth_range_general} reads
\begin{equation}
\lambda_{{\rm Fifth }0} = \frac{\sqrt{|f_{R0}|(n+1)}}{H_0}\left(\frac{\Omega_{m0} + 4 \Omega_{\Lambda 0}}{\Delta_\rho + 4 \Omega_{\Lambda 0}}\right)^{1 + \frac{n}{2}}
\end{equation}
which goes towards zero at $\Delta_\rho\equiv\rho_m/\rho_{c0}\rightarrow \infty$ while at the background level ($\Delta_\rho = \Omega_{m0}$) we have $\lambda_{{\rm Fifth }0} \simeq 3\sqrt{n+1} \sqrt{|f_{R0}| / 10^{-6}} \text{Mpc}\,h^{-1}$. This means, the higher the local density, the smaller the range of the fifth force and, thus, the previously described Chameleon screening is active.

\subsection{The $N$-body code \& simulation parameters}
The simulation code used is a modification of \texttt{RAMSES} \citep{Teyssier2002} which features the possibility of the inclusion of a scalar field. For a full description of the \texttt{ISIS} code see \citet{Llinares2014a}.

For each simulation run we used $512^3$ dark matter particles in a box with side-length $256\,{\rm Mpc}\,h^{-1}$. The cosmological parameters used are $(\Omega_{m0}, \Omega_{\Lambda 0}, H_0)=(0.267,\,0.733,\,71.9\,\mathrm{km}\,\mathrm{s}^{-1}\,\mathrm{Mpc}^{-1})$. These values correspond to a particle mass of $9.26\times 10^9\,M_{\sun} h^{-1}$. 

We selected three sets of parameters for the $f(R)$ model and four for the Symmetron model. All the model parameters are presented in Tab.~\ref{tab:run_params}. In addition, we ran one $\Lambda$CDM simulation which we will use as a reference when comparing the results. All simulations were performed using the same initial conditions generated using the \textit{Cosmics} package by \citet{Bertschinger1995astro.ph..6070B}.

\subsection{Halo selection}
\label{sec:halo-selection}
We identified cluster of galaxies using the `friends-of-friends` (FOF) halo finder \texttt{Rockstar} \citep{Behroozi2013}. Furthermore, we adopt the phase-space position of the halos given by \texttt{Rockstar}\footnote{This differs from the treatment in \citet{Gronke2014} where we \textit{merged} subhalos into it's mother-halo.}. 

Afterwards, we discard the unvirialized halos because we are interested in observables of dynamically relaxed halos rather than, e.g., major mergers. To do this we adapt the measure $\beta_{\rm vir}$ from \citet{Shaw2006}:
\begin{equation}
  \beta_{\rm vir}\equiv \frac{2 T - E_s}{W} + 1.
\end{equation}
Here, $T$, $W$ and $E_S$ are the kinetic, potential and surface pressure energies, respectively. We point out that it is crucial to use the particles' accelerations in order to calculate $W$ since otherwise the energy of the scalar field is neglected \citep[see][for details]{Gronke2014}.

A halo is defined to be sufficiently virialized if $|\beta_{\rm vir}|<0.2$. This cutoff means that $44$ percent of halos with mass $>10^{13}\,M_\odot h^{-1}$ have been discarded in the $\Lambda$CDM data set leaving a total of $4539$.

In addition, we used a second quantity in order to characterize the halos is the environmental variable $D_{\rm env}$ \citep{Haas2012MNRAS.419.2133H,Zhao2011a,Winther2011}. This variable which characterizes the environment of a halo is defined as 
\begin{equation}
D\equiv \frac{d_{M'\ge M}}{r_{M'\ge M}}.
\end{equation}
Here, $d_{M>M'}$ is the distance to the closest halo which has equal or greater mass than the halo of interest and $r_{M'\ge M}$ is the radius of that halo.
This is useful in the context of screened modified gravity models since a halo can not only be screened due to its own high density (\textit{self-screening}) but also through another nearby halo (\textit{environmental screening}). Following \citet{Zhao2011a,Winther2011}, we define a halo to be located in an \textit{overdense} environment if $D<10$. Otherwise ($D\ge 10$) we describe its surrounding as \textit{under dense}. We hope this helps us to disentangle the effects of environmental and self-screening.

\subsection{Analyzed quantities}
\label{sec:analyzed-quantities}

We will first present the global statistical properties of the velocity field that are described by the velocity divergence power spectrum.  Normalizing the divergence of the velocity field $\vec{\nabla}\cdot \mathbf{v}$ with the Hubble parameter gives the dimensionless expansion scalar
\begin{align}
\theta = \frac{1}{H}\vec{\nabla}\cdot \mathbf{v}.  
\end{align}
We compute the power-spectrum of $\theta$ from our simulations by using the public available {\tt DTFE} code \citep{2011arXiv1105.0370C} which calculates the Delaunay tessellation of the simulation volume. The velocity divergence field obtained this way gives us a field that is volume averaged rather than mass averaged. This yields an unbiased estimate for the power-spectrum which has better noise properties and is less sensitive to the details of the power-spectrum estimation (e.g., the grid-size used, the amount of empty cells, etc.). The velocity divergence is not the easiest quantity to measure in practice, but it is a useful diagnostic that can tell us to what degree the velocity field is affected by modified gravity. The spectra we will display are only shown up to half the Nyquist frequency $k_{\rm Nyq/2} \simeq 3$ Mpc$/h$ of the domain grid used in our simulations (which coincides with half the particle Nyquist frequency) to avoid resolution issues biasing our results \citep{2013MNRAS.428..743L}.

Besides that, our main interest is the velocity field inside the dark matter halos.  To characterize this quantity, we study 
the relative particle velocity, which is simply defined as
\begin{equation}
v_{\rm rel} = \sqrt{(\mathbf{v} - \mathbf{v}_{\mathcal H})^2}, 
\end{equation}
where $\mathbf{v}$ and $\mathbf{v}_{\mathcal H}$ are the particles' velocity and its halo velocity, respectively. For the latter, we use the core velocity of the halo, i.e., the mean particles' velocity within $10\%$ of the virial radius since this is expected to track the halo motion best \citep{Behroozi2013}.

Furthermore, we investigate the quantity $v_{\rm max}$ which is defined to be the maximum relative particle velocity within $R_{200}$. 

\begin{figure*}
  \centering
  \includegraphics[width=\linewidth]{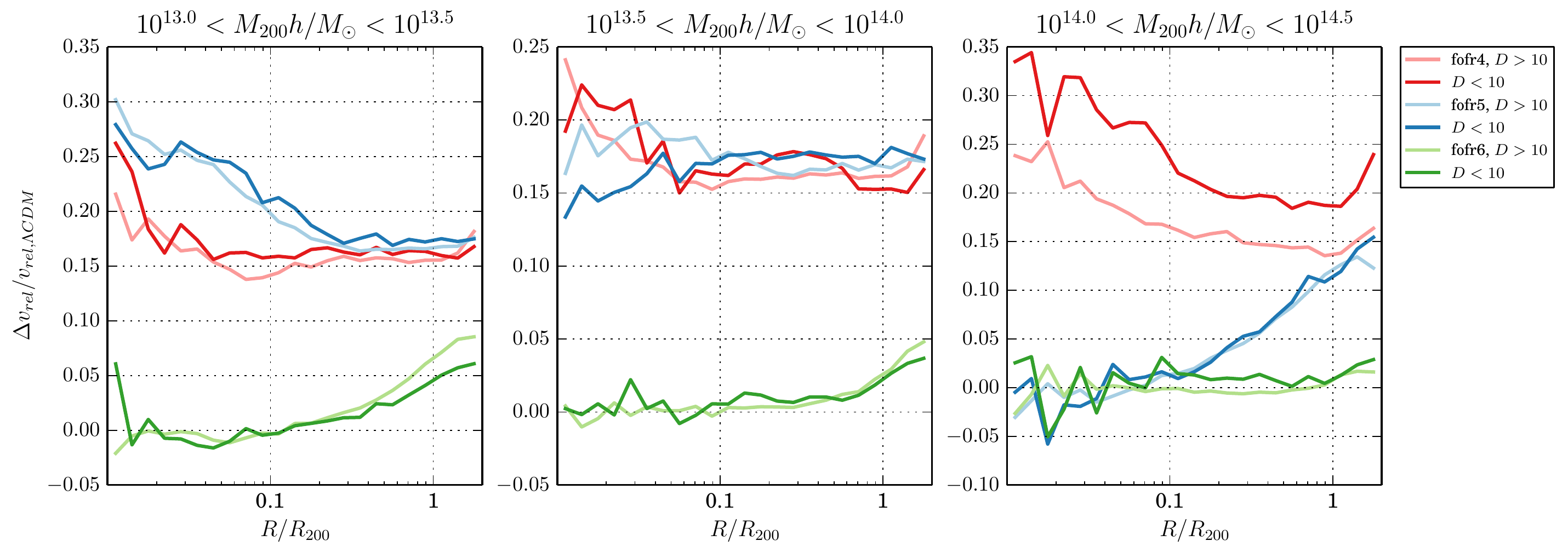}
  \caption{The relative particle velocity $\Delta v$ versus the normalized halo radius for the $f(R)$ model. The lighter (darker) curves show the values for halos residing in under(over)dense regions.}
  \label{fig:Dvrel_vs_R_fofr}
\end{figure*}

\begin{figure*}
  \centering
  \includegraphics[width=\linewidth]{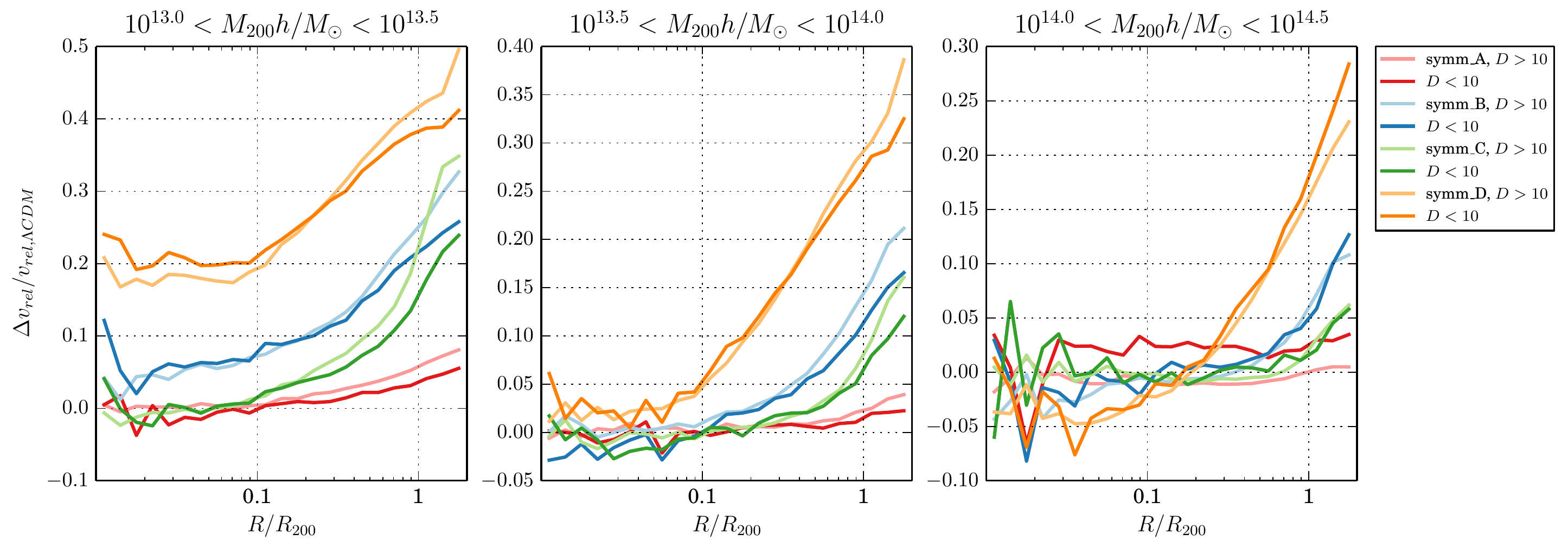}
  \caption{Same as \fig{fig:Dvrel_vs_R_fofr} but for the Symmetron.}
  \label{fig:Dvrel_vs_R_Symmetron}
\end{figure*}

\section{Results}
\label{sec:results}
When presenting our results we often choose to display the relative deviation to $\Lambda$CDM. That is for a quantity $X$, we show $(X_{\rm mg} - X_{\rm \Lambda\text{CDM}}) / X_{\rm \Lambda\text{CDM}}$. Here, $X_{\rm mg}$ and $X_{\rm \Lambda\text{CDM}}$ is the quantity measured in our modified gravity and $\Lambda$CDM simulations, respectively.

\subsection{Velocity divergence power-spectrum}
\label{sec:vel_div_pow_spec}
In \fig{fig:pofkv} we show the fractional difference in the velocity divergence with respect to $\Lambda$CDM for both modified gravity models. For comparison we also show the matter power-spectrum. 

For our $f(R)$ simulations (right panel in Fig.~\ref{fig:pofkv}) we find that the difference with respect to $\Lambda$CDM in the velocity divergence spectrum can be roughly two times as large as the difference in the matter power spectrum. These results agree very well with the findings of \citet{2013MNRAS.428..743L}. 

For the Symmetron -- displayed in the left panel of Fig.~\ref{fig:pofkv} -- the difference can be much larger. For the symm$\_$C model (which is the model with the largest value of the coupling strength $\beta$), we see that $(\Delta P/P)_{\rm m}\approx 10\%$ at $k=1h/\text{Mpc}$ while $(\Delta P/P)_{\theta}\approx 200\%$. The \symmC model has a fifth force in unscreened regions that is four times that of the other Symmetron models and this is likely the reason why we get this extreme signal. 

\subsection{Halo velocity profiles}
\label{sec:halo-veloc-prof}

We present The stacked halo velocity profiles in  in \figs{fig:Dvrel_vs_R_fofr}{fig:Dvrel_vs_R_Symmetron} for $f(R)$ and the Symmetron models, respectively. We show these halo profiles in three mass ranges, $[13, 13.5, 14]< \log_{10}\left(M_{200} h/M_\odot\right) < [13.5, 14, 14.5]$ and differ between halos residing in overdense ($D<10$) and underdense ($D>10$) regions.

For the $f(R)$ model (\fig{fig:Dvrel_vs_R_fofr}) we find the velocities boosted by $\sim 20\%$ in the \fofrfour case and by $\lesssim 5\%$ in the \fofrsix case for all the three halo mass ranges analyzed. Only for the \fofrfive parameters we find a mixed behaviour, that is, for the lower two halo mass ranges the boost is $\sim 20\%$ but for the heaviest halos there is no deviation from ${\rm \Lambda}$CDM in the inner parts and $\sim 15\%$ higher velocities are found in the outer parts.

This is different for the Symmetron results (displayed in \fig{fig:Dvrel_vs_R_Symmetron}). Here, we find mainly the contrary, i.e., the velocity boost in the outskirts is significantly different to the one found in the inner regions. This effect is particularly strong for the \symmD model but is also apparent (mostly for the smaller halo masses) for the \symmB and \symmC model. The \symmA model shows hardly any deviation from ${\rm \Lambda}$CDM (velocity boosts $\lesssim 5\%$).

For all the halo velocity profiles shown, the environment does not seem to play a major role since the low and high density curves follow the same trend. An exception is the highest mass bin of the \fofrfour case where some velocity offset can be found. However, since here the low density curve is actually showing less deviation from $\Lambda$CDM although the forces are not screened (see \fig{fig:gamma_vs_R_fofr}), this offset is likely due to statistical variations.

\subsection{Maximum cluster velocity}
\label{sec:maxim-clust-veloc}

\begin{figure}
  \centering
  \includegraphics[width=\linewidth]{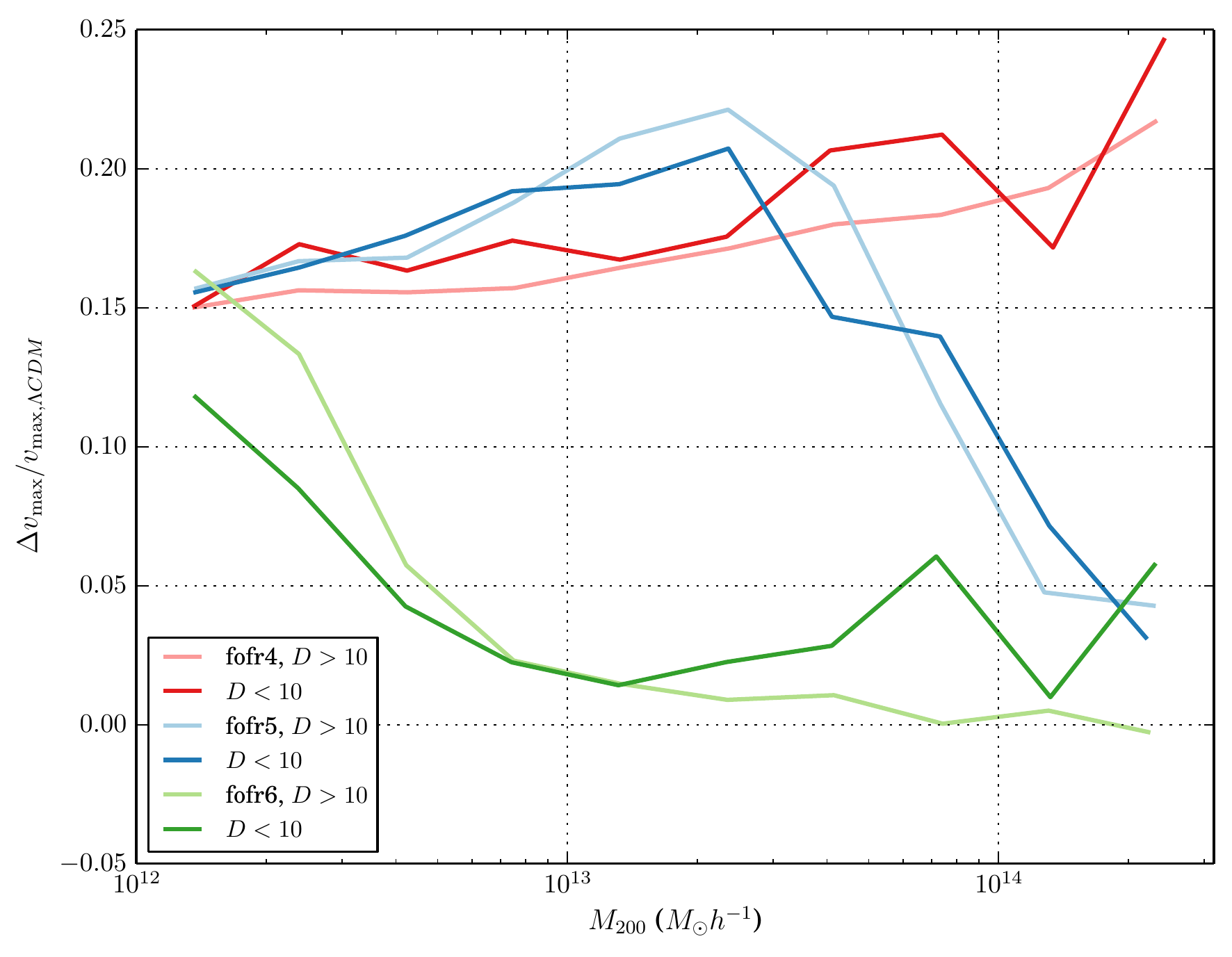}
  \caption{The relative deviation in the maximum cluster velocity as defined in Sec.~\ref{sec:analyzed-quantities} for the $f(R)$ model. Light and dark curves represent clusters located in under- and overdense regions, respectively.}
  \label{fig:Dvmax_vs_M_fofr}
\end{figure}

In \figs{fig:Dvmax_vs_M_fofr}{fig:Dvmax_vs_M_Symmetron} we show the deviation from $\Lambda$CDM of the maximum relative velocity found in a halo. This means, they provide another angle to the velocity-radius-mass relation considered in this paper.

\fig{fig:Dvmax_vs_M_fofr} shows the $f(R)$ results. For the \fofrfour model $v_{\rm max}$ is enhanced by $15\--20\%$ over the entire mass range. The \fofrfive model shows an comparable boost for masses $M_{200}\lesssim 2\times 10^{13}M_\odot h^{-1}$ but then the deviation drops sharply to $\lesssim 5\%$ at $2\times 10^{14}M_\odot h^{-1}$. There is indication that the same happens for the \fofrsix model but for a smaller cutoff mass. However, to reach a firm conclusion more simulations with a better mass resolution are necessary in order to resolve halos $M_{200}\lesssim 10^{12}M_\odot h^{-1}$.

\begin{figure}
  \centering
  \includegraphics[width=\linewidth]{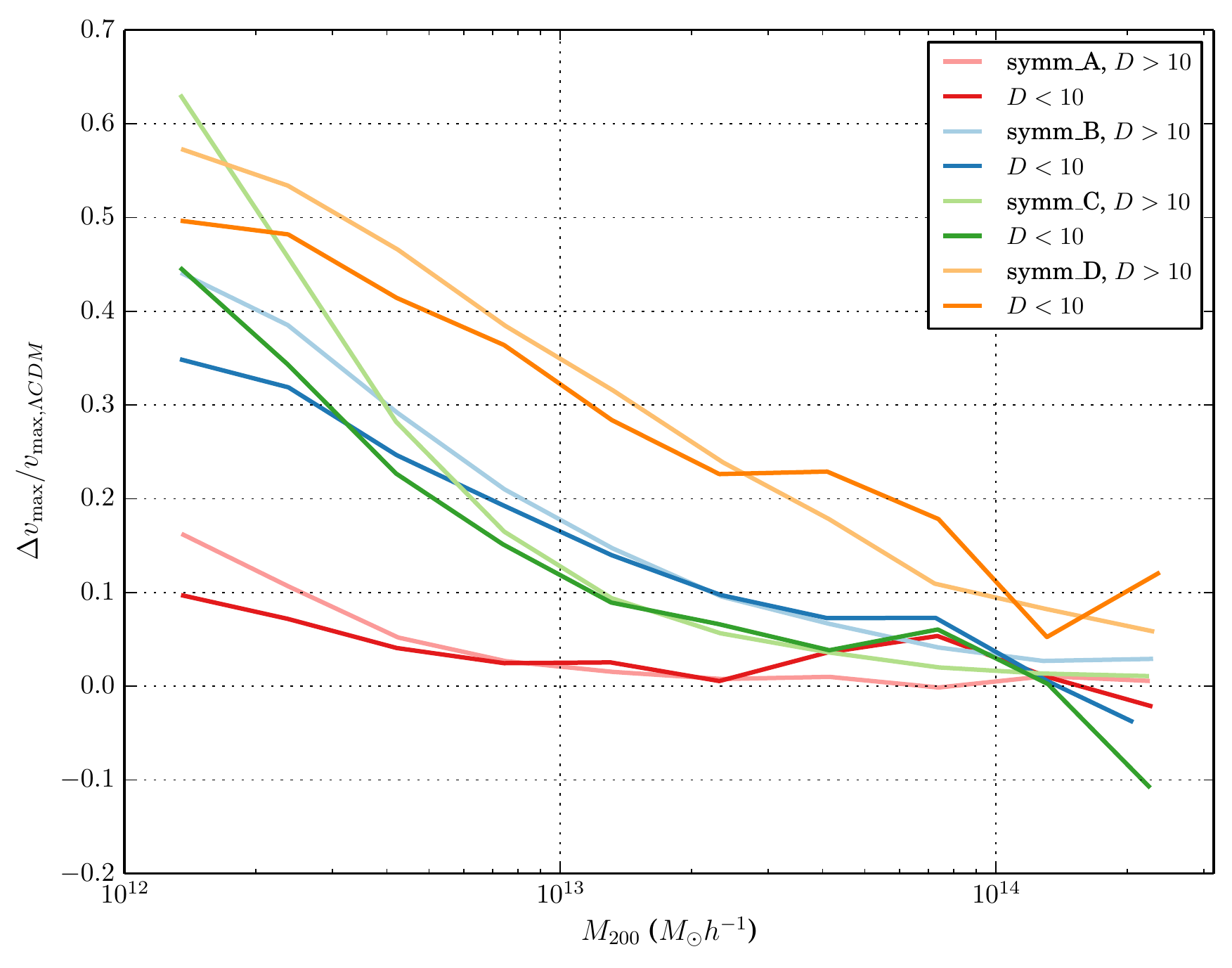}
  \caption{Same as in \fig{fig:Dvmax_vs_M_fofr} but for the Symmetron models.}
  \label{fig:Dvmax_vs_M_Symmetron}
\end{figure}

We show the same quantity, $\Delta v_{\rm max}/v_{{\rm max}, \Lambda\text{CDM}}$, for the Symmetron model in \fig{fig:Dvmax_vs_M_Symmetron}. Instead of a sharp cutoff we find a more gradual decline with halo mass. Overall, the $v_{\rm max}$ deviation is larger which is expected from the $\Delta v$ results. 

In both cases, we do find hardly any deviation which can be attributed clearly to environmental effects. The only exception are the Symmetron and \fofrsix results for the smallest ($12 \lesssim \log_{10}M_{200}h/M_\odot \lesssim 13$) where the deviation for halos within underdense regions is always larger than in overdense regions.

\subsection{Fifth force profiles}
\label{sec:fifth force-profiles}

\begin{figure*}
  \centering
  \includegraphics[width=\linewidth]{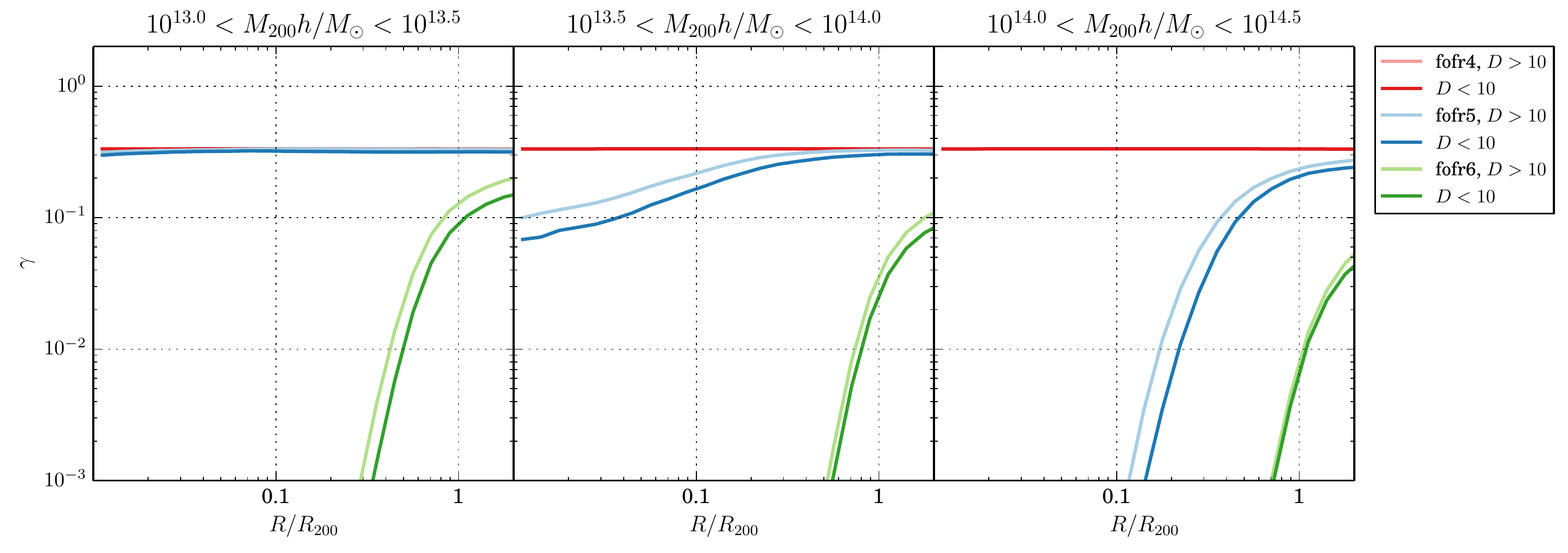}
  \caption{$\gamma\equiv |{\bf F_{\rm Fifth}}|/|{\bf F_N}|$ halo profiles for the $f(R)$ models studied. The light curves show halos where $D>10$, i.e. in underdense regions whereas the darker curves show halos with $D<10$.}
  \label{fig:gamma_vs_R_fofr}
\end{figure*}

\begin{figure*}
  \centering
  \includegraphics[width=\linewidth]{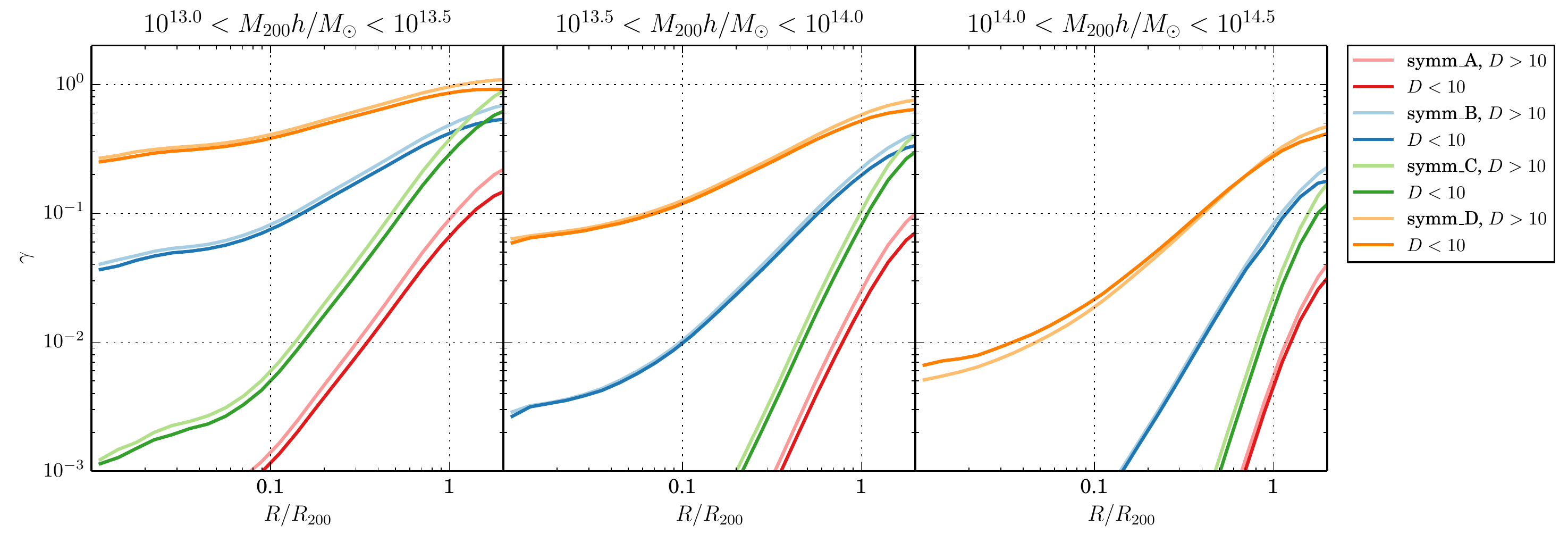}
  \caption{$\gamma\equiv |{\bf F_{\rm Fifth}}|/|{\bf F_N}|$ versus $R/R_{200}$ for the Symmetron models. As in \fig{fig:gamma_vs_R_fofr} indicate the lighter (darker) curves under(over)dense regions.}
  \label{fig:gamma_vs_R_Symmetron}
\end{figure*}

In order to understand the results obtained for the velocity distributions, we studied the force profiles inside the dark matter halos.  \figs{fig:gamma_vs_R_fofr}{fig:gamma_vs_R_Symmetron} show the quantity $\gamma\equiv |{\bf F_{\rm Fifth}}|/|{\bf F_N}|$ versus $R/R_{200}$ in the three previously used halo mass ranges for the $f(R)$ and the Symmetron model, respectively. 

The $f(R)$ results (\fig{fig:gamma_vs_R_fofr}) show that for the \fofrfour parameters, $\gamma$ is always at its theoretical maximum of $2\beta^2 = 1/3$ \citep{Hu2007} for all radii and all halos. This means that the density even within the heaviest halos considered is not sufficient to screen the fifth force. 
This looks different for the \fofrfive parameter set. Here, the force is completely screened in the center of the heaviest halos ($\gamma<10^{-3}$ for $R/R_{200}<0.1$) but not in outskirts. The medium-sized halos ($13.5 < \log_{10}\left(M_{200} h/M_\odot\right) < 14$) show also partial screening but not as strong since the density is not high enough to screen the fifth force in the same extend. The lightest halos are not screened in the \fofrfive case and, thus, $\gamma\approx 1/3$.
The simulation results with $f_{R0}=10^{-6}$ (the \fofrsix model) draw a completely different picture. In this case, all halo masses and radii show signs of screening. The strength of the screening is stronger for heavier halos and smaller radii as there the matter density is higher. All our $f(R)$ $\gamma$-profiles indicate that the environment plays some role as $\gamma(R, D<10) \le \gamma(R, D>10)$ but not a very strong one since the over- and underdense curves follow each other closely.

\fig{fig:gamma_vs_R_Symmetron} shows the $\gamma$-profiles for the four Symmetron parameter sets considered. Although the general behaviour is the same as for the $f(R)$ results, that is less fifth force in the center (for bigger halos) compared to the outskirts (to lighter halos), several differences are apparent. First, the $\gamma$ can reach higher values ($>1/3$) and no limiting value seems to be reached. Also, the form of screening differs. The Symmetron results are much more governed by \textit{partial screening}, that is where the fifth force is significant ($\gtrapprox 10\%$ of the Newtonian force) in the outskirts of the halo but screened ($\gamma\lessapprox 0.01$) in the inner parts.

Generally, we find in the Symmetron models that $\gamma$ is largest in the \symmD case, followed (in that order) by \symmB, \symmC and \symmD. This is true for all halo mass ranges and radii considered except for $R\approx R_{200}$ where the \symmC model might possess a stronger fifth force than the \symmB model.

Also for the Symmetron model we find that the environment plays a rather small role but has some effect leading to $\gamma(R, D<10) \le \gamma(R, D>10)$. However, for the largest halos this relation being inverted in one case (\symmD).

\section{Discussion}
\label{sec:discussion}

\subsection{Global velocity properties}
In order to put halo velocity profiles into perspective, we first studied the velocity divergence power spectrum in \S~\ref{sec:vel_div_pow_spec}. The velocity field in modified gravity simulations is found to be more affected by the presence of the fifth force than the density field as was noted previously by \citet{2013MNRAS.428..743L} for $f(R)$-gravity. For the Symmetron model we found this to be even more apparent. A particular striking example of this -- with boosts of up to $(\Delta P/P)_{\theta}\gtrsim 3$, whereas $(\Delta P/P)_{m}\sim 0.1$ -- are our \symmC results. Here, the fact that the \symmC model possesses a maximum fifth force which is four times that value in the other Symmetron models is likely the reason for obtaining such an extreme signal. Support for this explanation can be seen in \fig{fig:Dvmax_vs_M_Symmetron} where the maximum cluster velocity for small mass halos (which are more numerous in low-density regions) becomes greatest for the \symmC model.

The reason why generally, we have $(\Delta P/P)_{\theta}\gtrsim(\Delta P/P)_{m}$, is that the velocity divergence field is not mass-weighted in any way. Hence, low-density regions (voids) will contribute a large part of the signal in the velocity divergence power-spectrum (since voids contribute a large part of the volume in the Universe) which is not the case for the matter power-spectrum. Now the fifth-force is generally not screened in low-density regions so consequently velocities are boosted to significantly higher relative values (when compared with $\Lambda$CDM) in voids opposed to in clusters.
This indicates that low-density regions like cosmic voids, as we would expect, is the place where the strongest signals of modified gravity can be found\footnote{For studies of voids in modified gravity see, e.g., \citet{llinares_thesis,2012MNRAS.421.3481L, 2014arXiv1410.1510C}}.

\subsection{Halo velocity statistics}
Overall, we found that halo velocity profiles are an excellent direct trace of the fifth force: large values of $\gamma$ at a certain halo mass and radius, lead to large deviations in the relative velocity of particles at that point (see \S~\ref{sec:halo-veloc-prof} \& \S~\ref{sec:fifth force-profiles}). Hereby, we could not detect any differences due to the halo environment. This suggests that for most halos considered are well within the self-screening mass range \citep{Winther2011}, i.e., their own matter density dominates over that of the environment. This picture is supported by the fact that the only consistent discrepancy between the two environments can be found in the deviation of the maximum cluster velocity for $M_{200} \lesssim 10^{13} M_\odot h^{-1}$ in the Symmetron and the \fofrsix results. Here, the halos residing in overdense regions show less deviations from $\Lambda$CDM compared to ones in underdense regions. This might indicate the transition to the \textit{environmental-screening} mass range.

The exact impact of varying the specific model parameters (e.g., $a_{\rm ssb}$, $L$ and $\beta$ for the Symmetron) have already been discussed in detail in \citet{Gronke2014} and will not be repeated here. Instead, we want to focus on the categorization of the effects one can encounter within screened modified gravity theories. Namely, our findings suggest that one can group them into three general categories:
\begin{enumerate}
\item Fully screened regime. In this regime, the deviation in the relative particle velocities to $\Lambda$CDM is negligible for all halo radii. This is because here the fifth force is fully screened ($\gamma \ll 1$) everywhere. The \fofrsix model showed this behaviour for all considered halo masses as well as the \symmA, \symmB model for the bigger halos.
\item Unscreened regime. Here, the theoretical maximum of the strength of the fifth force is reached within the halo with the corresponding uniform particle velocity boost. An example of this are the \fofrfour results where $\gamma \sim 1/3$ and, thus, $\Delta v_{\rm rel}/v_{\rm rel, \Lambda\text{CDM}}\sim 0.2$ for all the considered halo masses ($13 \lesssim \log_{10} M_{200} h / M_\odot \lesssim 14.5$) and radii ($R < R_{200}$). The same is found in the \fofrfive case for halos in the mass range  $\log_{10} M_{200} h / M_\odot \lesssim 14$. On the other hand, none of the considered Symmetron models and halo masses shows a comparable behaviour.
\item Partially screened regime. All other velocity profiles show a large deviation from $\Lambda$CDM in the halo outskirts and less (none) in the central region of the halo. Here, the fifth force drops distinctly with decreasing halo radius (and, hence, matter density). A particular example of this is the \symmD model for all halo masses but also the other Symmetron parameter sets show a similar behaviour (at least for the smaller halo masses). Also, the \fofrfive results for the largest halo masses can be put within this category.
\end{enumerate}

These three regimes can be pointed out very clearly by analyzing the deviation in the maximum cluster velocity (see Sec.~\S~\ref{sec:maxim-clust-veloc}). Here, the fully screened regime corresponds to  $\Delta v_{\rm max}/v_{\rm max, \Lambda\text{CDM}}\sim 0$, the unscreened regime to a constant upper limit in deviation, i.e., to $Y\equiv \Delta v_{\rm max}/v_{\rm max, \Lambda\text{CDM}} > 0$ with $\dd Y/\dd M_{200} \sim 0$, and, the partially screened regime to the slope between (i) and (ii) (i.e., $Y > 0$ with $|\dd Y/\dd M_{200}| > 0$).
This confirms the previously mentioned points. All considered halos in the \fofrfour case are in the unscreened regime and in the \fofrfive case halos with $M_{200} \lesssim 3\times 10^{13} M_\odot h^{-1}$. To the contrary, the halos with $M_{200} \gtrsim 10^{13} M_\odot h^{-1}$ are in the fully screened regime in the \fofrsix and \symmD parameters. For the other Symmetron models studied all the considered halos are in the partially screened regime. 

Conclusively, we want to highlight the \fofrfive curves in \fig{fig:Dvmax_vs_M_fofr} which display nicely the transition from unscreened over partially screened to fully screened with increasing halo mass. A similar behaviour can be expected for all the considered screened modified gravity models. However, larger halo masses as well as a better resolution for smaller halos are needed in order to study this effect in more detail.

\section{Conclusions}
\label{sec:conclusions}
In this work, we studied the effect of screened-modified gravity theories on halo velocity profiles. To do this, we used the $N$-body code \texttt{ISIS} \citep{Llinares2014a} which includes the \citet{Hu2007} $f(R)$ as well as the Symmetron \citep{Hinterbichler} model and performed a set of $8$ simulations with $512^3$ particles. Our analyzed quantities were the power spectrum computed over the divergence of the velocity field, the relative particle velocity and its maximum value within a cluster.

Although the theoretical nature of the two screening mechanisms is completely different (see \S~\ref{sec:mod_grav_models}), we found common features in the velocity properties. In particular, we can highlight three distinct regimes: \textit{(i)} the fully screened regime where GR is recovered, \textit{(ii)} an unscreened regime where the strength of the fifth force is large, and, \textit{(iii)} a partially screened regime where screening occurs in the inner part of a halo, but the fifth force is active at larger radii.

Observationally, the partially screened regime might be of particular interest because the uniform increase of the gravitational force can also be due to a too low halo mass estimate. \citet{Zhao2011a,2010PhRvD..81j3002S} suggest therefore measuring lensing as well as dynamical masses since the former is not affected by many modified gravity theories. However, this is not always possible and, thus, searching for deviations from $\Lambda$CDM in the outskirts and the center of halos of a particular mass bin (i.e., the partially screened regime) might be a good alternative. It is noteworthy, that the deviations from $\Lambda$CDM are over-predicted in a CDM only simulations like ours \citep{Arnold2014MNRAS.440..833A,Puchwein2013MNRAS.436..348P,Hammami2014} and in order to constrain the parameter space of screened modified gravity theories like the Symmetron and Chameleon $f(R)$ models more detailed, hydrodynamical simulations are necessary. 

\section*{Acknowledgments}
CLL and DFM acknowledge support from the Research Council of Norway through grant $216756$. HAW is supported by the BIPAC and the Oxford Martin School. The simulations were performed on the NOTUR Clusters \texttt{HEXAGON}, the computing facilities at the Universities of Bergen, Norway.

\bibliography{references}

\label{lastpage}
\end{document}